%% file: beatrr.tex
\def\M{{\cal M}}

\def\FigCMD{{Figure 1}}
\def\FigRatio{{Figure 2}}
\def\FigPeter{{Figure 3}}
\def\FigDered{{Figure 4}}

\documentstyle[12pt,aasms4]{article}

\tighten

\begin{document}
\title{The MACHO Project LMC Variable Star Inventory IV:
Multimode RR Lyrae Stars, Distance to the LMC and Age of the Oldest Stars}
\author{
        C.~Alcock\altaffilmark{1,2}, 
      R.A.~Allsman\altaffilmark{3},
        D.~Alves\altaffilmark{1,2,4},
      T.S.~Axelrod\altaffilmark{5},
      A.C.~Becker\altaffilmark{6},
      D.P.~Bennett\altaffilmark{1,2,4,7},
      K.H.~Cook\altaffilmark{1,2},
      K.C.~Freeman\altaffilmark{5},
        K.~Griest\altaffilmark{2,8},
        J.~Guern\altaffilmark{2,8},
      M.J.~Lehner\altaffilmark{2,8},
      S.L.~Marshall\altaffilmark{1,2},
      D.~Minniti\altaffilmark{1,2},
      B.A.~Peterson\altaffilmark{5},
      M.R.~Pratt\altaffilmark{2,6,9},
      P.J.~Quinn\altaffilmark{10},
      A.W.~Rodgers\altaffilmark{5},
        W.~Sutherland\altaffilmark{11},
      D.L.~Welch\altaffilmark{12} \\
      {\bf (The MACHO Collaboration) }
      }
\input institutions.tex

\begin{abstract}

We report the discovery of 73 double-mode RR Lyrae (RRd) stars in fields 
near the bar of the LMC. The stars are detected among the MACHO database 
of short-period variables that currently contains about 7900 RR Lyrae stars. 
Fundamental periods (P$_0$) for these stars are found 
in the range 0.46-0.55 days and first overtone-to-fundamental
period ratios are found to be in the range $0.742< P_1/P_0 < 0.748$. 
A significant fraction of our current sample have period ratios smaller
than any previously discovered RRd variables. We present mean
magnitudes, colors, and lightcurve properties for all LMC RRd stars
detected to date. The range in period ratios is unexpectedly large.

We present a determination of absolute magnitudes for these stars
based primarily on pulsation theory and the assumption that all observed
stars are at the fundamental blue edge (FBE) of the instability strip.
Comparison of the calibrated MACHO V and R$_{KC}$ photometry with these 
derived absolute magnitudes yields an absorption-corrected distance modulus 
to the LMC of $18.57 \pm 0.19$ mag which is in good agreement with 
that found (18.5) through comparison of galactic and LMC Cepheids. 

Adopting this luminosity calibration, we derive an increase in the 
distance modulus, and thus a reduction in the age found via isochrone fitting
for M15 of about 33\% and discuss the implications for cosmology.

\end{abstract}

\section{Introduction}

Multimode RR Lyrae, also known as RRd type, are especially
interesting tests of pulsation and evolutionary theory because
precise period ratios allow an estimate of the mass of each
star. This paper reports properties of a total of 73 such
stars in the Large Magellanic Cloud (LMC), nearly twice as large
as the number of all known RRd stars prior to this work.

The MACHO Project LMC RR Lyrae database is described 
in Alcock {\it et al.} (1996) where accumulated V and R photometry for 
approximately 7900 RR Lyrae stars
in 22 intensively-observed fields has been analysed to determine
the distributions of period and amplitude. In that paper, a random
selection of 500 lightcurves was examined and two multimode stars were 
identified. The ratios of first overtone-to-fundamental
period, $P_1$/$P_0$, of those stars were found to be lower than 
any previously discovered RRd stars.  

This paper reports the discovery of a further 71 RRd stars. 
We first discuss the observational properties of this new
sample. Next we estimate absolute magnitudes using pulsation theory.
Finally, we note implications of the derived luminosities for 
cosmology.

\section{Results}

\subsection{Photometry}

The photometric reduction procedures for the 
MACHO images are described in Alcock {\it et al.} (1996). 
The MACHO instrumental magnitudes have been placed on the standard 
photometric system of Kron-Cousins V and R through comparison 
with \markcite{Landolt1992} Landolt (1992) standard stars that were 
transferred to the 22 high-priority MACHO fields.  
While gross zero-point differences between
fields have been removed, small offsets within each
field exist as an artifact of the MACHO photometry reduction procedure.
We conservatively adopt an error in V or R of $\pm$0.10 mag.
The transformation between MACHO instrumental color and Kron-Cousins (V-R)
is less affected by the internal zero-point offsets, therefore, the color
for each star in Table 1 has an estimated uncertainty of $\pm$0.05 mag.
We arrive at these estimates of the photometric uncertainties through
numerous consistency checks within the MACHO database and by comparisons
with published photometry of LMC stars.

The lightcurves used in this analysis represent all data acquired for 
these stars during an interval of approximately 1400 days and 
typically represent 400-700 points per color. In \FigCMD\ we show the
mean apparent V and (V-R) for LMC double-mode RR Lyrae and illustrate
their position relative to other LMC field stars.  The lightcurves
for these stars are available on the MACHO Project home pages
at the URLs {\tt http://wwwmacho.anu.edu.au/ and
http://wwwmacho.mcmaster.ca/} .

\subsection {Selection Criteria}

Double-mode RR Lyrae candidates were selected from our color-magnitude
diagram-based RR Lyrae sample by two techniques. The first, described in
\markcite{Alcock1996}
Alcock {\it et al.} (1996), relied on visual examination of 500 lightcurves
using the IRAF 
\footnote{Developed at the National Optical Astronomical Observatories}
PDM task. More recently, candidate RRd stars were selected
based on their measure-of-fit to a singly-periodic lightcurve, their period
and lightcurve amplitude. 

The strong period dependence of our selection criteria deserves mention.
For photometric periods between 0.35 and 0.40 days -
corresponding to fundamental periods between 0.46 and 0.50 days - 
our selection criteria identify double-mode RR Lyrae very efficiently.
Shortward and longward of this range, our yield is reduced by the 
increase in false positives with period estimates of one-third and
one-half of a day, respectively. The measure-of-fit statistic is
less effective at long periods due to the presence of Blazhko
variation in some RRab pulsators. Therefore, we caution the reader that
our current sample is not complete and that the distribution in
period (and therefore mass) is unrepresentative.

\subsection {Observed Properties of the Sample}

We summarize our discovery of LMC RRd stars in Table 1. This table lists,
from left-to-right, the equinox J2000.0 right ascension and declination,
the internal database identifier, the number of measurements and mean
magnitude in V and R$_{KC}$, respectively, the estimate of the fundamental
period $P_0$ (in days) and the period ratio, $P_1$/$P_0$, the semi-amplitude
of the Fourier component for the frequency corresponding to $P_1$ (in mags),
and the ratio of the amplitudes for the first Fourier components of the
first-overtone and fundamental modes. The uncertainties in the period and
period ratio are expressed in parentheses in units of the least significant
digits. Table 1 contains 75 entries. The pairs ({\tt 13.6691.4052},
{\tt 6.6691.1003}) and ({\tt 13.6810.2845}, {\tt 6.6810.428}) represent 
the same stars found twice in overlapping regions of different MACHO fields. 
The total number of distinct stars is therefore 73. We remind the reader
that our preferred nomenclature is MACHO* followed by the right ascension
and declination as listed in Table 1, with no spaces.

The periods of the RR Lyrae were initially estimated using the 
IRAF PDM task as described in Alcock {\it et al.} (1996). 
The periods, period
ratios and modal amplitudes reported in Table 1 are the result of a
weighted least-squares fit to the RRd lightcurves for the first three
Fourier terms in each mode and all coupling terms.

The ratio of first-overtone to fundamental mode amplitudes is 
shown in \FigRatio. It is clear that this sample of stars is similar
to known RRd's in that, almost without exception, the amplitude of
the first-overtone mode is greater than that of the fundamental. We
note that the selection criteria used above are biased against stars
with fundamental periods shorter than 0.46 days and that amplitude
ratios smaller than unity may therefore be under-represented.

\subsection {Absolute Magnitudes from Pulsation Theory}

It is, in theory, possible to estimate absolute magnitudes of RRd stars
by combining the Stefan-Boltzmann equation and the period-density 
relations. The result of such an exercise is an expression relating 
period to luminosity, effective temperature, and mass. Unlike singly-periodic
RR Lyrae stars, the period ratio allows us to estimate the mass based
on atmospheric pulsation models. Provided that we can determine the
effective temperature, the set of required observables is complete.
Such a relationship is well-known for Cepheid variables where the
existence of the instability strip boundaries gives rise to an
obvious period-luminosity relation. For double-mode RR Lyrae stars
we expect a similar bounded region in the color-magnitude diagram
which will also give rise to a period-luminosity relationship.

We will now proceed through the steps necessary to obtain absolute
magnitude estimates. First, the estimation of masses from periods and
period ratios will be described. Second, the estimation of effective
temperature and the relationship of these stars with the fundamental
blue edge (FBE) of the instability strip will be described. Third,
we substitute for both effective temperature and mass in the pulsation
equation and produce a period-luminosity relation. Finally, we estimate
the errors introduced at each step.

\subsubsection {Period Ratios and Masses}

We show in \FigPeter, the Petersen (1973) diagram of $P_1$/$P_0$ 
plotted against $P_0$ for stars in M15 (Nemec, 1995a), 
Draco (Nemec, 1995b), 
IC 4499 (Clement {\it et al.}, 1986, Walker and Nemec, 1996) and the LMC. 
We additionally mark on the diagram the locus of 
period ratios for model pulsators computed by Cox (1991) using OPAL91 
opacities.

The distribution of the stars in \FigPeter\ shows that the LMC stars 
include a low mass extension to the multimode stars discovered in other 
systems. From \FigPeter\ it can be shown that a relation of the form
\begin{equation}
\log P_{0} = 0.452 \log \M - 0.228            
\end{equation}
applies across the indicated mass range of the multimode stars in the LMC 
and the Galactic Oosterhoff groups I and II. 

The shortest period of Bailey type a,b stars among the total 
sample of LMC RR Lyrae stars is 0.459 days and, additionally, with a period 
ratio of about 0.7421 the overtone periods are close to one-third of a day 
and are thus more difficult to discover due to aliasing. 
We mark these period regions as hatched 
areas in \FigPeter\ to emphasize the degree to which the 
period-distribution of the multimode stars is affected by selection 
biases in this work.
Nevertheless, it is clear that the tight locus of observed period
and period-ratio in \FigPeter\ for RRd stars born in a wide range of
environments is consistent with a tight relationship between
luminosity, effective temperature and mass.
In the following discussion we shall assign masses to the LMC stars 
on the basis of their position in the Petersen diagram relative to the 
Cox (1991) models. We note that there are potential uncertainties in these 
masses due to the lack of specific data on composition 
of the LMC stars. However. the abundance distribution of the LMC field RR 
Lyrae stars as derived from spectra does not deviate strongly from that 
of the galactic halo (\markcite{Alcock1996} Alcock {\it et al.} 1996). 

\subsubsection{Pulsation Models}

The pulsation equation for model envelopes of RR Lyrae stars
has been characterized using OPAL95 opacities 
by \markcite{Bono1996} Bono {\it et al.} (1996) as
\begin{equation}
\log P_{0,tr} = 11.627 + 0.823 \log L_{tr} - 3.506 \log T_{e,tr} - 0.582 \log \M_{tr}  
\end{equation}
over the mass range 0.65-0.75 \M$_{\odot}$.
We have annotated the variables to reflect the assumption that the 
multimode stars are in the transition zone (the region between 
the fundamental blue edge and the first-overtone red edge) 
of the instability strip. 
Their equation 2 is a linear fit to the results of period determinations 
for RR Lyrae star models using the most recent composition parameters 
in a study of the non-linear pulsation properties. The fit to the 
pulsation equation is in the spirit of the influential study of RR Lyrae 
star pulsation by \markcite{vAB1971} van Albada and Baker (1971) 
and we note in passing that use of their form of the pulsation 
equation does not affect the results and conclusions drawn in this paper.

We make the further assumption (see, for example, 
\markcite{Caputo1990} Caputo 1990, 
\markcite{Bono1994} Bono and Stellingwerf 1994) 
that there is close similarity between the 
temperature of the blue edge, (FBE) of the fundamental instability strip 
and the transition zone occupied by the multimode stars. 
$P_{tr}$ is thus taken as the period of the fundamental mode at 
the blue edge of the instability strip. In the range of $P_{tr}$ of the 
beat RR Lyrae stars, $T_{e,tr}$ varies from 6900 $K$ to 7060 $K$ 
in excellent agreement with the calculations of the fundamental and 
first-overtone instability strip boundaries (and the zone where both 
fundamental and first overtone are unstable), by 
\markcite{Bono1994} Bono and Stellingwerf (1994). 
Their calculations show that the temperature of the strip in 
which pulsation in either the fundamental or first-overtone 
mode is allowed is relatively insensitive to mass or luminosity and
is centered near 6840 K for the mass range 0.58-0.75 \M$_{\odot}$
and luminosity range 1.6-1.9 $\log L_{\odot}$.
The assumption that the double-mode stars 
lie near this temperature is also supported by the close 
similarity of the periods of the shortest period LMC Bailey 
type a,b stars at $P_0$ = 0.457 days to the average period of the 
multimode stars at $P_0$ = 0.48 days.

For the FBE, we can derive a relationship between the fundamental 
period and the effective temperature by combining equation 8 of
\markcite{Sandage1993a}
Sandage (1993a) -- the relation between the shortest period for
RRab stars and [Fe/H] -- and equation 5 of 
\markcite{Sandage1993b} Sandage (1993b) 
-- the relation between $\log T_e$ and [Fe/H] for RRab stars -- to obtain:
\begin{equation}
\log T_{e,tr} = 3.816 - 0.0984 \log P_{tr}.    
\end{equation}
We see that the FBE is very nearly vertical in the $\log L$, $\log T_e$ 
plane. Sandage expresses both the 
FBE period and temperature in terms of abundance [Fe/H] but 
since the range of [Fe/H] in the LMC field RR Lyrae stars is the 
same as in the calibrating galactic RR Lyrae clusters and field, 
we have written the FBE temperature directly in terms of period. 
The luminosity of the multimode stars is then
\begin{equation}
\log L_{tr} = 2.483 + 2.360 \log P_{tr}    
\end{equation}
Over the range of the transition periods covering the majority of the LMC 
multimode stars shown in \FigPeter, at $P_0$ = 0.46 days, $\log L_{tr}= 1.69$ 
and at $P_0$ = 0.52 days, $\log L_{tr} = 1.81$, corresponding to absolute 
visual magnitudes of $M_V$ = $+0.49$ and $+0.19$ respectively when the small 
bolometric corrections of 
\markcite{Kurucz1979} Kurucz (1979) are applied. For the longer 
fundamental periods of the multimode stars in the metal-weak galactic 
globular cluster M 15 with a characteristic $P_0$ = 0.54 days, the derived luminosity 
corresponds 
to $M_V$ = $+0.05$ mag. Most other calibrations of RR Lyrae star 
luminosities have been expressed in terms of composition and there 
has been considerable debate about the slope of the coefficient of 
the term in [Fe/H].  We note that if we adopt the composition value 
of [Fe/H] = -2.12 for M15 
(\markcite{Buonanno1989} Buonanno {\it et al.} 1989), our calibration has a 
roughly similar dependence on [Fe/H] to that discussed by 
\markcite{Sandage1993a} \markcite{Sandage1993b} \markcite{Sandage1993c} 
Sandage (1993a,b,c) but caution that 
direct comparison of these dependencies is limited by the small number 
of clusters that contain multimode stars. We also note that our 
calibration depends on the theoretical pulsation equation which, 
as we will show in the next section, appears to be supported by the 
luminosities of the LMC RR Lyrae stars. 
Ours is a brighter calibration of RR Lyrae star luminosities than 
many previously derived. This calibration receives support from 
the independent study of the Bailey type c RR Lyrae stars by 
Simon and Clement (1993) who compared the Fourier phase parameters 
and periods of the first overtone pulsators with linear and 
hydrodynamic pulsation models to determine temperatures and 
a luminosity calibration. The result of their calibration applied to 
the Reticulum cluster of the LMC yields roughly
the same distance modulus as we derive below for 
the field LMC stars. Their discussion of the galactic multimode 
stars as checks on their calibration of the RRc variables 
additionally supports the assumptions we made above concerning 
the location of the multimode instability strip in the 
luminosity-temperature plane.  

In his recent monograph, \markcite{Smith1995} Smith (1995) 
reviews at least 6 methods, both fully and semi-empirical, 
which yield luminosities roughly 0.4 mags fainter than those 
derived here. Conversely, all these methods would predict the 
LMC RR Lyrae stars to be too faint by the same amount. In this paper 
we adopt the view that the characteristics of RR Lyrae stars in the 
LMC and the Galaxy are properly comparable and explore the consequences 
of that comparison. 

\subsubsection{Error Analysis}

The application of the luminosity calibration derived in the previous
section to the photometry of our multimode RR Lyrae stars in the LMC 
necessitates a careful examination of the uncertainties.
We adopt a mass uncertainty of $\pm$0.05 $\M_{\odot}$ 
(\markcite{Kovacs1992} Kovacs {\it et al.} 1992) in our $\log P_0$, 
$\log \M$ relation. Assuming double-mode RR Lyrae stars lie at the FBE, 
the uncertainty in equation (3), our $\log T_{e,tr}$, $\log P_{tr}$ 
relation derived from 
\markcite{Sandage1993a}\markcite{Sandage1993b}
Sandage (1993a,b), is dominated by the uncertainty of the (B-V), 
$log T_{e,tr}$ relation used to 
define the temperatures of galactic RR Lyrae along the FBE.  
We return to the unpublished model atmosphere
calculations of Bell as seen in \markcite{Butler1978} Butler {\it et al.} 
(1978), the model atmospheres of \markcite{Kurucz1979} Kurucz (1979), 
and the instability strip analysis of \markcite{Bono1996}
Bono {\it et al.} (1996), and by intercomparison adopt at the FBE
an uncertainty of 0.01 in $\log T_{e,tr}$. 
Propagating the errors through to equation (4),
we find that at a given period the uncertainty in $\log L_{tr}$ is 
$\pm$0.065 resulting in an uncertainty in $M_V$ of $\pm$0.16 mag.
We have assumed equation (4) is an exact representation and there is
negligible error in these bolometric corrections.

\subsubsection{The LMC Distance Modulus}

In \FigCMD\ we show the multimode stars in the V,(V-R) color-magnitude 
diagram (CMD). We adopt the reddening vector with $\rm R_V = A_V/E(V-R) 
= 5.0$ (\markcite{Bessel1996} Bessell, 1996) which is illustrated in
\FigCMD. Also shown is the adopted photometric uncertainty for each 
point. The insert shows the multimode stars in 
relation to a representative MACHO LMC CMD.
Using the temperatures from equation (3), for stars with 
$\log g = 2.6$, we adopt the (B-V), (V-R) relation 
(\markcite{Bessel1996} Bessell 1996),
\begin{equation}
(V-R) = 0.004 + 0.566(B-V)
\end{equation}
with an uncertainty of $\pm$ 0.017 in the zero point. 
From \markcite{Butler1978} Butler {\it et al.} (1978) we interpolate  
the T (our adopted temperature for the FBE), (B-V) relations between 
[Fe/H] = -1 and -2 for [Fe/H] = -1.7, a value appropriate for the LMC 
RR Lyrae (\markcite{Alcock1996} Alcock {\it et al.} 1996) and 
derive a $(V-R)_{0}$, $\log P_{tr}$ relation,
\begin{equation}
(V-R)_0 = 0.19 + 0.15\log P_{tr}
\end{equation}
with an uncertainty in the zero-point of $\pm$0.02 mag derived
from the uncertainties in equations (3) and (5) as discussed
previously.  Adopting our FBE temperature, the intrinsic color,
$(V-R)_0$, ranges from 0.14 to 0.15 mag for our multimode stars. 
Furthermore, we note that the mean reddening, $<E(V-R)>$ = 0.049 mag, 
or $<E(B-V)>$ = 0.086 mag, agrees well with other measurements of the 
reddening toward the LMC (\markcite{Bessell1991} Bessell 1991).
In \FigDered\ we show the de-reddened V
magnitudes plotted against $P_0$.  The error bars noted are derived
from the uncertainty in V and in (V-R) scaled by the ratio of total
to selective absorption ($R_V$ = 5), equivalent
to $\pm$0.27 mag.  The contribution of the uncertainty in our adopted
intrinsic color is systematic and we have not included it in the
uncertainty of each individual de-reddened magnitude. Our multimode 
RR Lyrae have a mean period $<P{_0}>$ = 0.482 corresponding to 
$<M_{V}>$ = 0.37 $\pm$ 0.16 mag. The mean de-reddened magnitude of 
our multimode RR Lyrae sample is $<V_{0}>$ =  18.94 $\pm$ 0.03 mag, 
in excellent agreement with the mean de-reddened V magnitude of 
nearly 180 RR Lyrae found in LMC clusters (\markcite{Walker1992} 
Walker 1992). The distance modulus of the LMC naturally
follows: $(m-M)_{LMC}$ = 18.57 $\pm$ 0.19 mag, where the final error
is a combination of the uncertainty in $(V-R)_0$ (scaled
by $R_V$ = 5), the uncertainty in $<V_{0}>$, and the uncertainty
in $<M_{V}>$.  Fitting the data of \FigDered\ to the $M_{V}$, $\log P_{tr}$
calibration with a fixed slope yields the same LMC distance modulus.
The best fit distance modulus is shown as a horizontal
line in \FigDered. The corresponding $M_{V}$, $\log P_{e,tr}$, 
calibration is also shown.
The distance modulus for the LMC multimode RR Lyrae stars derived here is 
found to be consistent with that ($18.5 \pm 0.1$ mag) derived from the 
galactic calibration of classical Cepheids discussed by 
\markcite{Walker1992} Walker (1992).

\section{ Discussion }

\markcite{Walker1989}\markcite{Walker1992} Walker (1989,1992) has 
emphasized the apparent discrepancy of 0.3 magnitudes between 
RR Lyrae star luminosities derived from Baade-Wesselink analyses of 
galactic stars and the luminosities of LMC cluster variables if the 
Cepheid distance calibration is correct. We have shown here that 
there is a consistency between the luminosities derived from the pulsation
equation, and independent luminosity measures such as the 
LMC Cepheid distance calibration and the SN1987A ring distance
(\markcite{Gould1995} Gould 1995). Consequently,
the pulsation properties of the multimode RR Lyrae stars appear to
be a useful tool in distance studies through their calibration of 
the more common and easily discovered RRab stars.

The unusually large range in $P_1$/$P_0$ and the range in implied masses
appear to present new questions about the evolution from the ZAHB. 
The \markcite{Lee1990} Lee and Demarque (1990) and 
\markcite{Sweigert1987} Sweigert (1987) HB evolutionary tracks 
do not carry the evolution of stars with masses as low as 
$0.55M_\odot$ as far as the instability strip but it does not appear 
that such stars could be of similar luminosity to the HB when they 
reach the instability strip following redward evolution. 
\markcite{Cox1991} Cox (1991) goes further and
states that stars with mass of $0.55M_\odot$ would never reach the
instability strip at all. However as we see in \FigPeter, there
do exist LMC multimode RR Lyrae stars with this mass as derived from 
pulsation theory. In \markcite{Alcock1996} Alcock {\it et al.} (1996), 
we argued that the field HB of the LMC
has a red morphology and that as a result, the beat RR Lyrae
stars are rare. Our more recent detections still show fewer of these stars 
among the LMC field population than, say in the OoII cluster M15.  
We remind the reader that our survey is incomplete and that this fraction
may rise. In the same paper, we also argued that the LMC field 
RR Lyrae star abundance distribution was similar to the halo galactic 
distribution with a modal ${\rm[Fe/H]} = -1.7$. These findings led us to 
conclude that the LMC field was younger than most galactic globular 
clusters. 

An implication of \FigDered\ and the period-luminosity relation  contained 
in equation (4) is a differential rescaling of the distances to 
galactic globular clusters (and the galactocentric distance) when 
HB luminosities are the criteria. Oosterhoff group II clusters are 
the most affected. 

\markcite{Walker1992} Walker (1992), in an important paper, 
supposed that the distance to the LMC was correctly 
defined by the galactic Cepheid distance scale and inferred that 
the LMC RR Lyrae stars were 0.3 mag brighter than previously suggested. 
He examined the consequences of a global 
increase in HB star luminosities on the age and distance of galactic 
globular clusters. In this paper we have determined the RR Lyrae star 
luminosities to be significantly brighter than previously thought 
and have shown that this 
revision is most important for the OoII clusters in the Galaxy. 
For illustrative purposes we discuss the archetypal Galactic cluster M15 
which has multimode RR Lyrae stars with a mean fundamental period of 
$P_0$ = 0.54 days. Equation (4) implies the HB $M_V = +0.05$ mag.   
\markcite{Buonanno1989} Buonanno {\it et al.} (1989) list the 
brightness difference between MSTO and HB as 3.54 mag for this cluster, 
leading to $M_{V,to} = 3.59$ mag. These authors extract from evolutionary 
calculations the relation 
\begin {equation}
\log t_9 = -0.41 + 0.37M_{V,to} - 0.43Y - 0.13[Fe/H]
\end {equation}
between age, turn-off luminosity and composition. For M15 we take Y = 0.23 
and [Fe/H] = -2.15 (see \markcite{Buonanno1989} Buonanno 
{\it et al.} 1989)). This equation yields an evolutionary age of 
12.6 Gyr. (This estimate does not include the effect of any alpha-particle 
enrichment.) \markcite{Minniti1996} Minniti {\it et al.} (1996) have found
[O/Fe] = 0.45 for M15 and we estimate that this results in a reduction in age 
of 0.6 Gyr based on the work of \markcite{Salaris1993} 
Salaris {\it et al.} (1993).
The age estimate of 12.0 Gyr has the characteristic uncertainty of 
$\pm$1.5 Gyr associated with MSTO ages. These age estimates can 
undoubtedly be refined by the fitting of evolutionary tracks to 
the recalibrated CMD's. We note that compared to a HB absolute magnitude
of $M_V$ = +0.49 mag for M15 (\markcite{Armandroff1989} Armandroff 1989)
our calibration produces a reduction in age of 33\% which is relatively 
independent of the assumed value of alpha-element enrichment.

A fuller exploration of the parameter space of cosmological models defined by 
the age of the oldest stars is beyond the scope of this paper but we note 
that a 33\% reduction in the estimated age of the oldest stars will help 
reconcile globular cluster ages with recent estimates of the Hubble 
constant, $H_0$.

\acknowledgements

We are grateful for the skilled support given our project by the
technical staff at Mt. Stromlo Observatory (MSO). Work performed at 
Lawrence Livermore National Laboratory (LLNL)
is supported by the Department of Energy (DOE) under contract 
W7405-ENG-48. Work performed by the Center for Particle Astrophysics 
(CfPA) on the University of California campuses is supported in part 
by the Office of Science and Technology Centers of the National 
Science Foundation (NSF) under cooperative agreement AST-8809616. 
Work performed at MSO is supported by the Bilateral Science and
Technology Program of the Australian Department of Industry,
Technology and Regional Development. KG acknowledges a DOE OJI grant,
and the support of the Sloan Foundation. DLW
was a Natural Sciences and Engineering Research Council (NSERC)
University Research Fellow during this work.

\newpage

\begin{figure}
\plotone{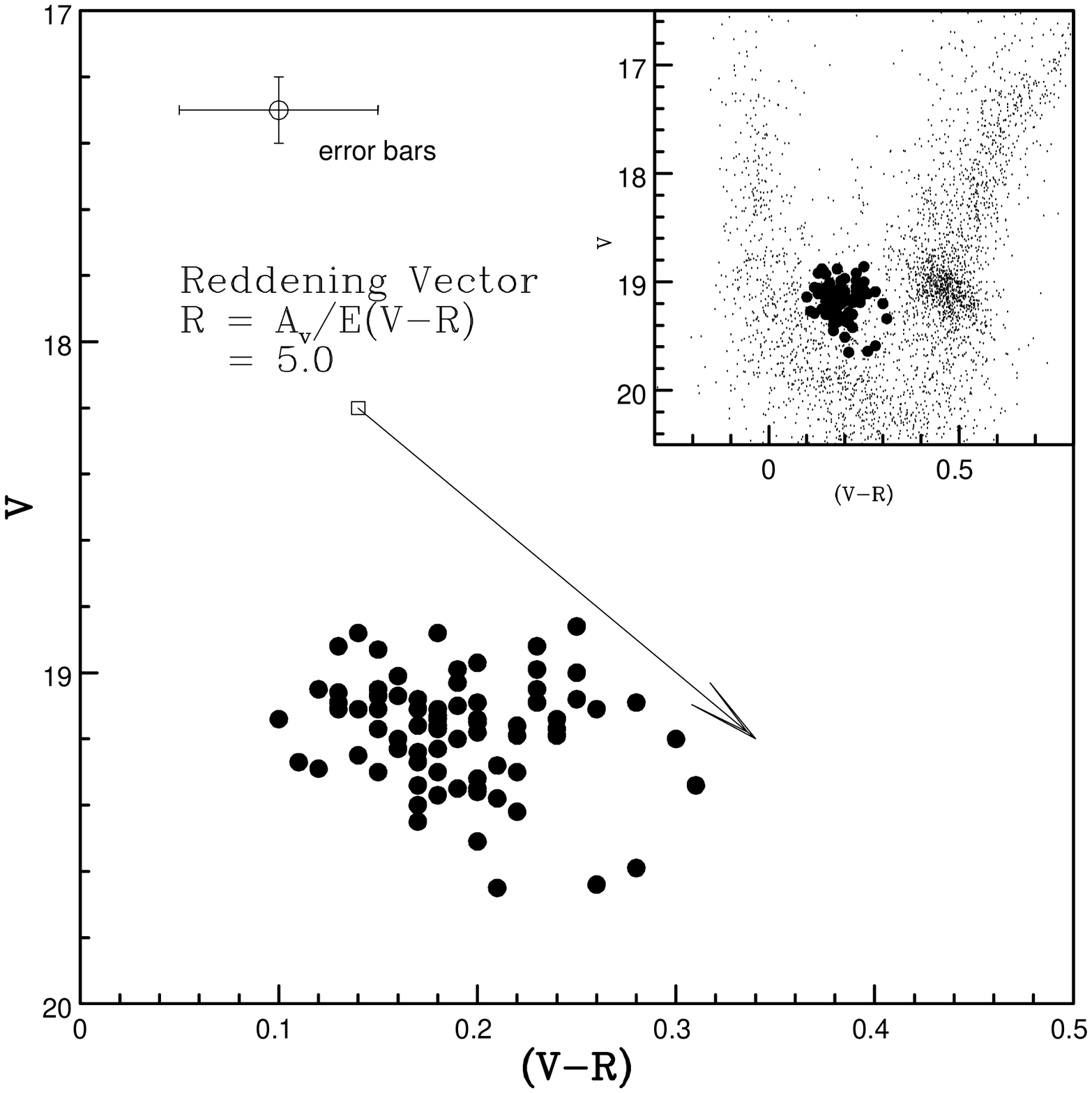}
\caption{Observed median V magnitudes are plotted against derived
(V-R) colors for the multimode RR Lyrae stars. The reddening vector
is marked with an arrow.
The insert shows the CMD of the multimode stars (filled circles), 
in relation to a representative CMD in an LMC field.}
\end{figure}

\begin{figure}
\plotone{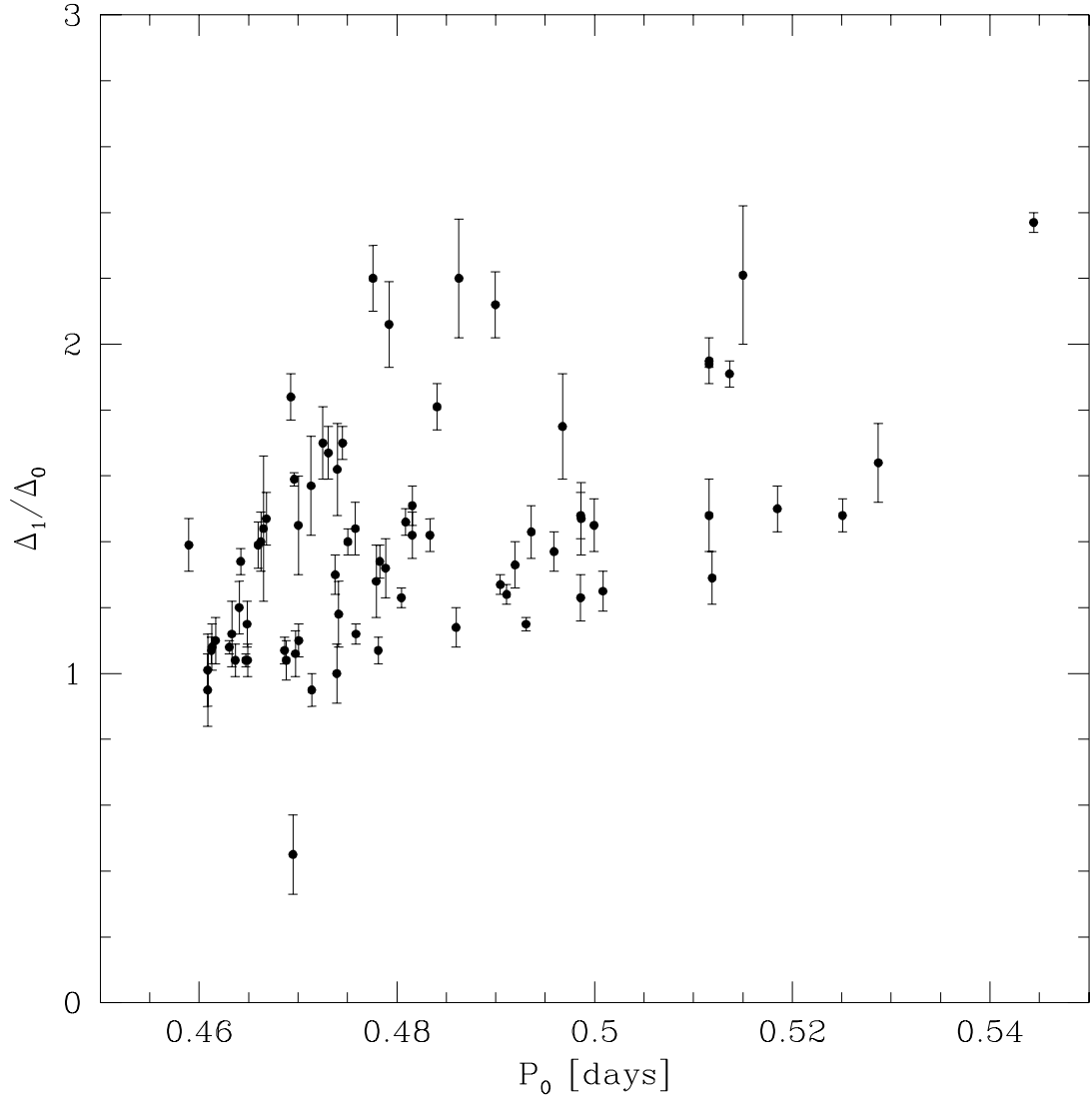}
\caption{The ratio of the amplitudes of the first Fourier components
for the first-overtone and fundamental modes, $\Delta_1$/$\Delta_0$,
are plotted against the fundamental mode period $P_0$. Almost without
exception, the amplitude of the first overtone mode is greater than
that of the fundamental - a property that this sample shares with
previously discovered RRd stars in globular clusters and dwarf
spheroidal galaxies. We caution, however, that our sample is
incomplete due to period-dependent selection biases and that more
RRd with $\Delta_1$/$\Delta_0 < 1$ may yet be discovered in our LMC
sample.}
\end{figure}

\begin{figure}
\plotone{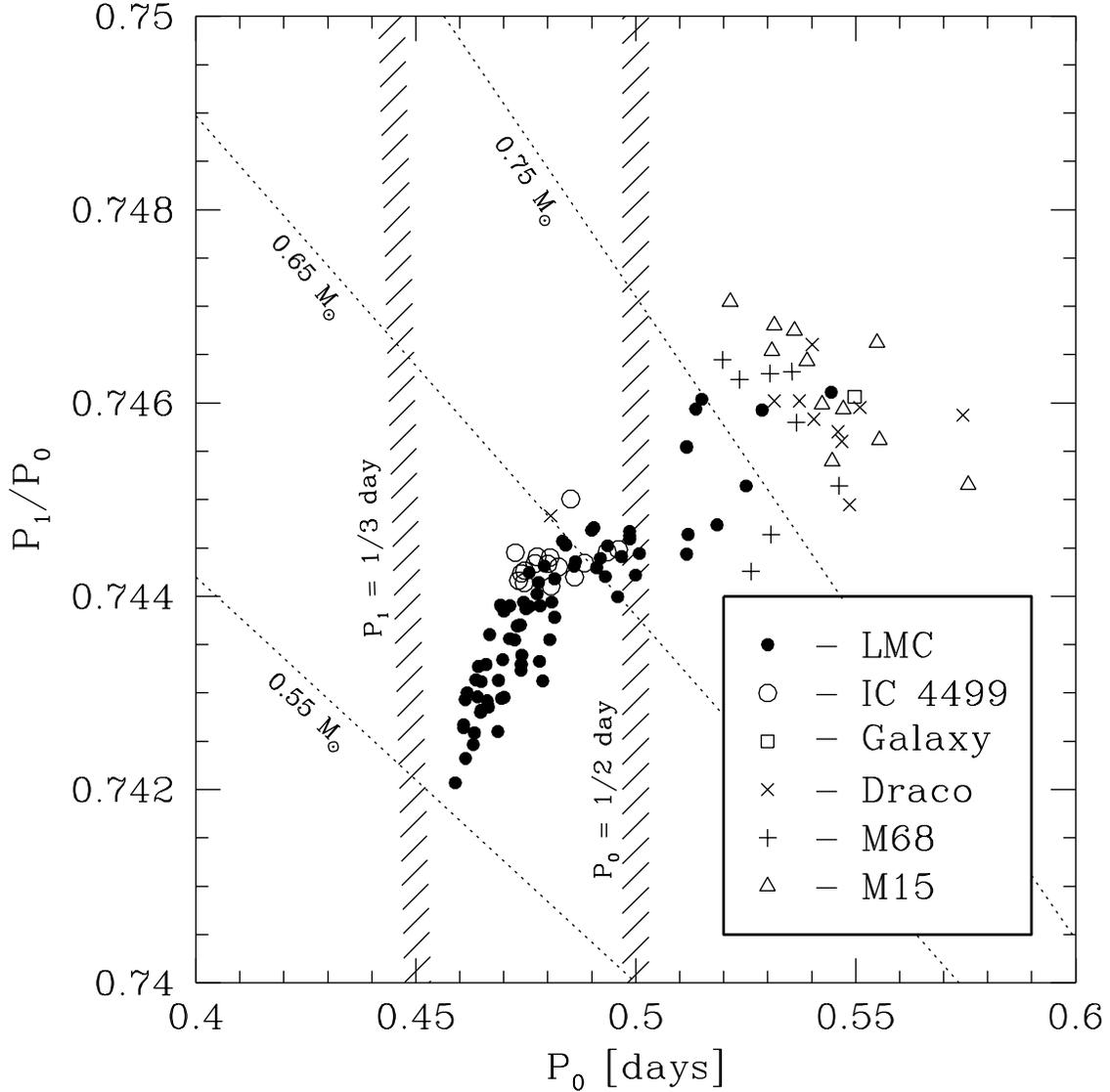}
\caption{The Petersen diagram for multimode RR Lyrae stars where the
first-overtone to fundamental period ratios $P_1$/$P_0$ are plotted against
$P_0$. As described in the text, the LMC stars all have $P_0$ $<$ 0.5 days
while the longer period stars are from M15 and Draco. The loci of 
model pulsators
calculated by Cox (1991) are shown for the masses of 0.55, 0.65 and 0.75
$\M_\odot$. The near-vertical line at $P_0$ = 0.46 days marks the location 
of the one-third day alias for the first-overtone period.}
\end{figure}

\begin{figure}
\plotone{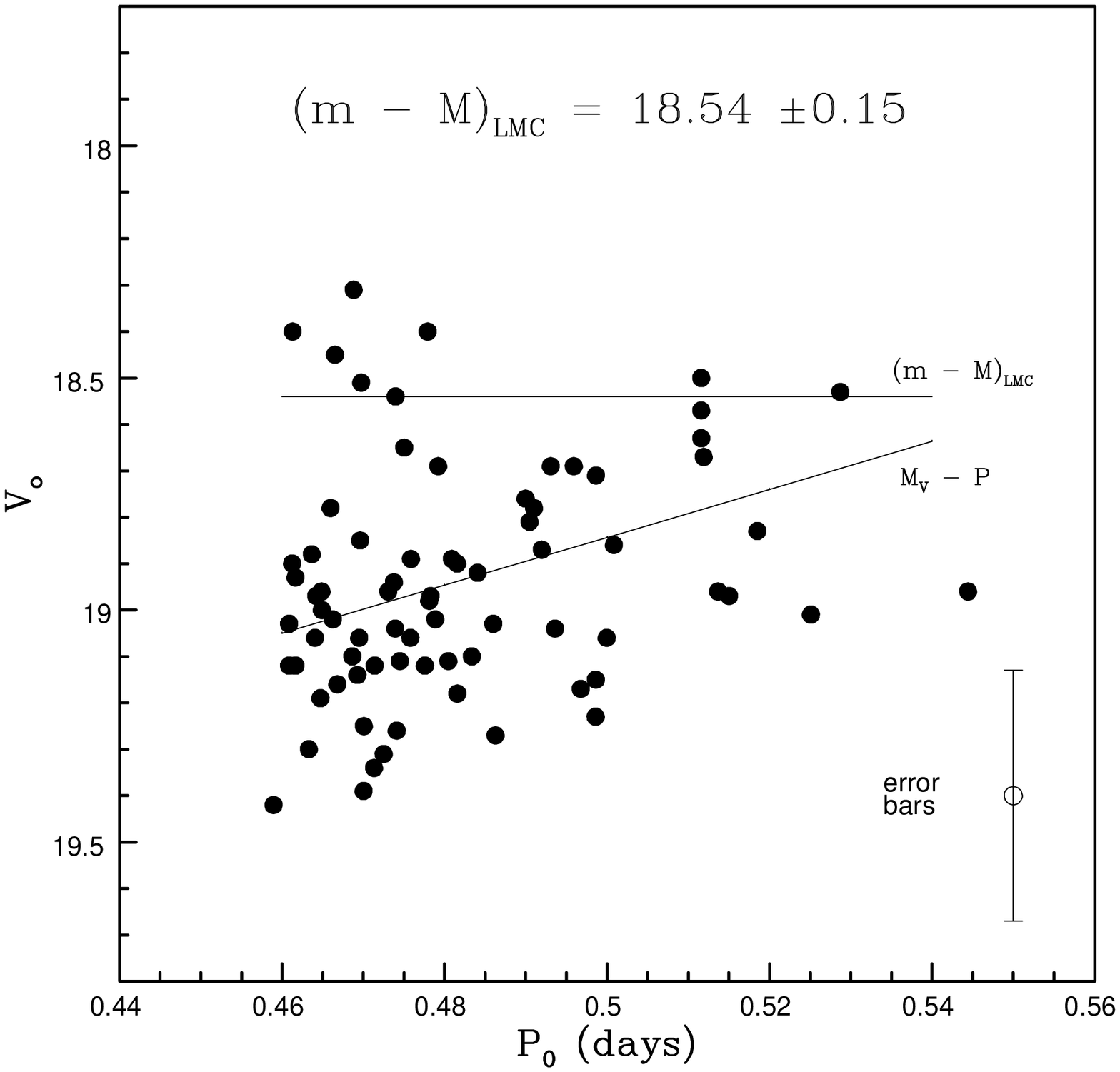}
\caption{The absorption-corrected magnitudes, $V_0$, are plotted 
against the fundamental period $P_0$. The line has the slope given 
by the equation (4).}
\end{figure}

\end{document}

%% file: institutions.tex
\altaffiltext{1}{Lawrence Livermore National Laboratory, Livermore, CA 94550\\
        E-mail: {\tt alcock, alves, bennett, dminniti, kcook, stuart@igpp.llnl.gov}}
 
\altaffiltext{2}{Center for Particle Astrophysics,
        University of California, Berkeley, CA 94720}
 
\altaffiltext{3}{Supercomputing Facility, Australian National University, Canberra, ACT 0200, \\
        Australia, E-mail: {\tt robyn@macho.anu.edu.au}}
 
\altaffiltext{4}{Department of Physics, University of California,
        Davis, CA 95616 }
 
\altaffiltext{5}{Mt.~Stromlo and Siding Spring Observatories, Australian National University,
        Weston \\ Creek, ACT 2611, Australia , E-mail: {\tt tsa, kcf, peterson, alex@mso.anu.edu.au}}
 
\altaffiltext{6}{Departments of Astronomy and Physics,
        University of Washington, Seattle, WA 98195\\
        E-mail: {\tt becker, mrp@astro.washington.edu}}

\altaffiltext{7}{Physics Department, University of Notre Dame, Notre Dame, IN 46556}
 
\altaffiltext{8}{Department of Physics, University of California,
        San Diego, La Jolla, CA 92093\\
        E-mail: {\tt kgriest, jguern, mlehner@ucsd.edu }}
 
\altaffiltext{9}{Department of Physics, University of California,
        Santa Barbara, CA 93106 }
 
\altaffiltext{10}{European Southern Observatory, Karl-Schwarzchild Str. 2,
        D-85748, Garching, Germany \\
        E-mail: {\tt pjq@eso.org}}

\altaffiltext{11}{Department of Physics, University of Oxford, Oxford OX1 3RH, U.K. \\
        E-mail: {\tt w.sutherland@physics.ox.ac.uk}}
 
\altaffiltext{12}{Dept. of Physics \& Astronomy, McMaster University, Hamilton, Ontario, \\
        L8S 4M1 Canada, E-mail: {\tt welch@physics.mcmaster.ca}}